\documentclass[12pt]{article}
\usepackage{amsmath}
\usepackage{amssymb}
\usepackage{graphicx}
\usepackage{ifthen}
\usepackage{verbatim}
\usepackage{psfrag}

\numberwithin{equation}{section}

\renewcommand{\theequation}{\arabic{equation}}

\begin{document}

\begin{flushright}
\texttt{hep-th/0611324}\\
\texttt{OU-HET 570}\\
November 2006
\end{flushright}
\bigskip
\bigskip
\begin{center}
{\Large \textbf{Plethystics and instantons on ALE spaces}}
\end{center}

\vspace{10mm}
\begin{center}
Yui Noma
\footnote{E-mail: \texttt{yuhii@het.phys.sci.osaka-u.ac.jp}},
Toshio Nakatsu
\footnote{E-mail: \texttt{nakatsu@het.phys.sci.osaka-u.ac.jp}}
and 
Takeshi Tamakoshi
\footnote{E-mail: \texttt{tamakoshi@het.phys.sci.osaka-u.ac.jp}}\\
\bigskip
{\small 
\textit{Department of Physics, Graduate School of Science, 
Osaka University,\\
Toyonaka, Osaka 560-0043, Japan\\}}
\end{center}

\vspace{10mm}
\begin{abstract}
We present an expression
 of a deformed partition function
 for $\mathcal{N}=2$ $U(1)$ gauge theory
 on $\mathbb{C}^2/\mathbb{Z}_k$
 by using
 plethystic exponentials.
\end{abstract}

\section*{}

Recently 
 $\mathcal{N}=2$ instantons on ALE spaces
 were studied in
 \cite{Fucito:2004ry,Fujii:2005dk}
 and applied to black hole physics in
 \cite{Griguolo:2006kp,Fucito:2006kn}.
In this short note 
 we point out an expression of a deformed partition function
 for $\mathcal{N}=2$ $U(1)$ gauge theory
 on $\mathbb{C}^2/\mathbb{Z}_k$
 by using plethystic exponentials.
Plethystic exponentials have been used
 in \cite{Benvenuti:2006qr}
 and play a central role in counting the BPS operators.

Let us start with
 describing the partition function for 
 $\mathcal{N}=2$ $U(1)$ gauge theory
 on $\mathbb{C}^2/\mathbb{Z}_k$.
The partition function
 was computed in \cite{Fujii:2005dk}
 and becomes as follows:
\begin{eqnarray}
&&
Z(\epsilon_1, \epsilon_2;\Lambda, k) 
=
 \sum_{\substack{\lambda \\ \lambda=\mu(\phi; \lambda^{(r)})}}
 \Lambda^{2 |\lambda|/k}
  \prod_{\substack{s\in \lambda \\ h(s)=0 \,\mathrm{mod} \,k}}
 \frac{1}{(\epsilon_1 (l(s)+1) -\epsilon_2 a(s))
 (- \epsilon_1 l(s) +\epsilon_2 (a(s)+1))}\,,
\nonumber \\ 
\label{eq;partition for 4d}
\end{eqnarray}
where $\epsilon_1, \epsilon_2$
 are parameters of
 the standard 
 torus action 
 on $\mathbb{C}^2/\mathbb{Z}_k$,
 and $\Lambda$ denotes
 a scale parameter for the gauge theory.
The sum in eq.(\ref{eq;partition for 4d}) 
 is a summation over partitions 
 (the Young diagrams). 
We remark that any partition $\lambda$ 
can be decomposed into a single partition 
$\lambda_{\mathrm{core}}$ and $k$-tuple partitions 
$\lambda^{(1)}, \cdots, \lambda^{(k)}$
\cite{James,olsson}. 
They are called respectively 
$k$-core and $k$-quotients. 
This arises from a division algorithm for 
the Young diagrams analogous to that for integers.
See Appendix \ref{sec;appendix} for details. 
We may write 
$\lambda=\mu(\lambda_{\mathrm{core}}; \lambda^{(r)})$ 
to emphasize the decomposition. 
The sum in eq.(\ref{eq;partition for 4d}) 
is restricted so that only partitions 
whose $k$-cores are empty  
contribute to the partition function. 
$\phi$ denotes the empty partition.
The symbol $|\lambda|$ is the number of boxes in the Young diagram: 
$|\lambda|:=\sum_{i=1}^{\infty}\lambda_{i}$. 
For a box $s$ in the Young diagram, 
$h(s)$, $a(s)$ and $l(s)$ are respectively, 
as given in (\ref{eq;arm and leg}),  
the hook length, the arm length and the leg length. 
In eq.(\ref{eq;partition for 4d}), 
the products are taken over boxes whose hook length 
are multiples of $k$.

We will deform the partition function (\ref{eq;partition for 4d}) 
by using a parameter $R$ in the following manner:
\begin{eqnarray}
Z(\epsilon_1, \epsilon_2;\Lambda,k, R)
=
 \sum_{\substack{\lambda \\ \lambda=\mu(\phi;\lambda^{(r)})}}
 \nu^{|\lambda|/k}
 \prod_{\substack{s\in \lambda \\ h(s)=0 \,\mathrm{mod} \,k}}
 \frac{1}{(1- t_1^{l(s)+1} t_2^{-a(s)})(1- t_1^{-l(s)} t_2^{a(s)+1})},
\label{eq;q-deformed partition function}
\end{eqnarray}
where
\begin{eqnarray}
t_1=e^{\epsilon_1 R},\hspace{5mm}
t_2=e^{\epsilon_2 R},\hspace{5mm}
\nu= (R\Lambda)^2.
\end{eqnarray} 
The parameter $R$ is identified 
in the gauge theory 
with a circumference of a circle in the fifth direction.  
For the case of $k=1$, 
the deformed partition function 
is calculated in \cite{Awata:2005fa} 
from the view point of refined topological vertices. 
It is clear that 
eq.(\ref{eq;q-deformed partition function}) 
reduces to eq.(\ref{eq;partition for 4d}) 
at the limit where $R$ goes to zero:
\begin{eqnarray}
\lim_{R \rightarrow 0}
 Z(\epsilon_1, \epsilon_2;\Lambda,k, R)
=
 Z(\epsilon_1, \epsilon_2;\Lambda, k).
\end{eqnarray}

Plethystic exponentials 
appeared in the study of counting the BPS operators 
\cite{Benvenuti:2006qr}.
In particular, by using it, 
counting functions of holomorphic functions (the BPS operators) 
over the symmetric products 
$S^N( \mathbb{C}^2)$, 
where the operators are counted 
weighted by their $U(1)$ $R$ charges 
(charges of the torus action), 
are reproduced from the counting function for $N=1$, 
that is, $\mathbb{C}^2$. 
For a given function $f(t_1,t_2)$,  
plethystic exponential $PE(f(t_1,t_2))$ is defined by
\begin{eqnarray}
PE\left(f(t_1,t_2)\right)
&:=&
 \exp\left(
\sum_{n=1}^{\infty}
 \frac{\nu^n}{n} f(t_1^n,t_2^n)
\right). 
\end{eqnarray}
See \cite{plethysm, plethystic exponential} 
for details of plethystic exponentials. 
We point out the following expression of 
the deformed partition function 
by means of plethystic exponentials: \\
\fbox{\begin{minipage}{170mm}
\begin{eqnarray}
Z(\epsilon_1, \epsilon_2;\Lambda,k, R)
 &=&
PE\left( h_k(t_1, t_2)\right),
\label{eq;main claim}
\end{eqnarray}
where
\begin{eqnarray}
h_k(t_1, t_2)
&:=&
\frac{1-t_1^k t_2^k}{(1-t_1^k)(1-t_2^k)(1-t_1 t_2)}.
\end{eqnarray}
\end{minipage}}
The RHS of (\ref{eq;main claim})
 can be simply written as 
\begin{eqnarray}
PE\left(h_k(t_1,t_2)\right)
=
 \prod_{\substack{m,\,n=0 \\ m=n\,\,\mathrm{mod}\, k}}^\infty
 \frac{1}{(1-\nu t_1^m t_2^n)}.
\end{eqnarray}
The function $h_k(t_1,t_2)$ is a counting function of 
holomorphic functions on $\mathbb{C}^2/\mathbb{Z}_k$. 
The ring of the holomorphic functions is 
generated by 
$X,Y$ and $Z$ subject to the relation $XY=Z^k$. 
The $U(1)$ charges are read as $(k,0)$ for $X$, 
$(0,k)$ for $Y$ and $(1,1)$ for $Z$. 
The equality (\ref{eq;main claim}) for the case of $k=1$ is 
due to the identities related with Macdonald's function
\cite{macdonald}. 
It is also understood 
as one parameter deformation of that 
made by Fujii and Minabe \cite{Fujii:2005dk}. 
By letting $R\rightarrow 0$, 
(\ref{eq;main claim}) reduces to their result:
\begin{eqnarray}
\lim_{R \rightarrow 0}
PE\left(h_k(t_1,t_2)\right)
=
\exp\left(
 \frac{\Lambda^2}{k\epsilon_1\epsilon_2}
\right)
=
Z(\epsilon_1, \epsilon_2;\Lambda, k).
\end{eqnarray}

\begin{figure}[htb]
\begin{center}
\psfrag{a1}{$\lambda^{(1)}=\phi$,}
\psfrag{a2}{$\lambda^{(2)}=\phi$,}
\psfrag{a3}{$\lambda^{(3)}=(2)$,}
\psfrag{b1}{$\lambda^{(1)}=\phi$,}
\psfrag{b2}{$\lambda^{(2)}=(1)$,}
\psfrag{b3}{$\lambda^{(3)}=(1)$,}
\psfrag{c1}{$\lambda^{(1)}=(1)$,}
\psfrag{c2}{$\lambda^{(2)}=\phi$,}
\psfrag{c3}{$\lambda^{(3)}=(1)$,}
\psfrag{d1}{$\lambda^{(1)}=\phi$,}
\psfrag{d2}{$\lambda^{(2)}=(2)$,}
\psfrag{d3}{$\lambda^{(3)}=\phi$,}
\psfrag{e1}{$\lambda^{(1)}=(1)$,}
\psfrag{e2}{$\lambda^{(2)}=(1)$,}
\psfrag{e3}{$\lambda^{(3)}=\phi$,}
\psfrag{f1}{$\lambda^{(1)}=(2)$,}
\psfrag{f2}{$\lambda^{(2)}=\phi$,}
\psfrag{f3}{$\lambda^{(3)}=\phi$,}
\psfrag{g1}{$\lambda^{(1)}=\phi$,}
\psfrag{g2}{$\lambda^{(2)}=\phi$,}
\psfrag{g3}{$\lambda^{(3)}=(1,1)$,}
\psfrag{h1}{$\lambda^{(1)}=\phi$,}
\psfrag{h2}{$\lambda^{(2)}=(1,1)$,}
\psfrag{h3}{$\lambda^{(3)}=\phi$,}
\psfrag{i1}{$\lambda^{(1)}=(1,1)$,}
\psfrag{i2}{$\lambda^{(2)}=\phi$,}
\psfrag{i3}{$\lambda^{(3)}=\phi$.}
\includegraphics[scale=1.0]{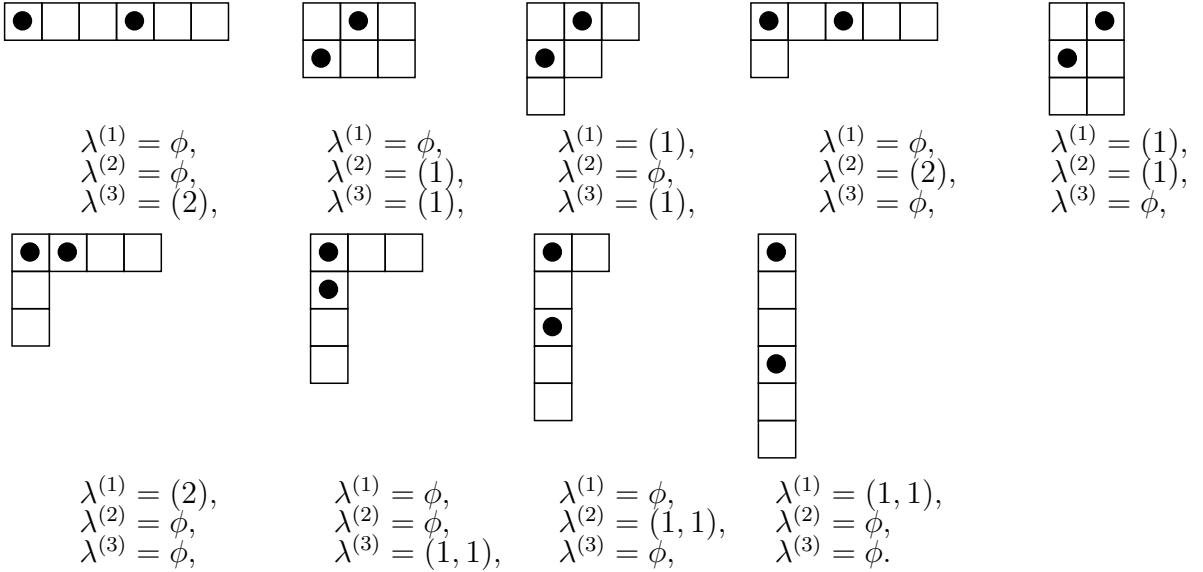}
\caption{{\it Young diagrams which 
 contribute to the coefficient of $\nu^2$ 
 in eq.{\rm (\ref{eq;q-deformed partition function})} 
 for the case of $k=3$.
The boxes filled with black circles 
 contribute to the products in 
 eq.{\rm (\ref{eq;q-deformed partition function})}.
$3$-tuple partitions attached
 below each Young diagram are the corresponding
 $3$-quotients.
}}
\label{fig;partitions_w_circle}
\end{center}
\end{figure}

We are convinced of the equality (\ref{eq;main claim}) 
by comparing the first few coefficients 
in the Taylor expansions with respect to $\nu$ 
of the both hand sides. 
As an example, let us see the case of $k=3$.
We will compare the coefficients of $\nu^2$. 
The Young diagrams which contribute to the coefficient 
of $\nu^2$ in (\ref{eq;q-deformed partition function}) 
are those of six boxes as shown 
in Figure \ref{fig;partitions_w_circle}.
In each Young diagram, 
only the boxes filled with black circles 
contribute to the products 
in (\ref{eq;q-deformed partition function}).
We then see 
\begin{eqnarray}
\mbox{Coeff}_{\nu^2}\left(
Z(\epsilon_1, \epsilon_2;\Lambda,3, R)
\right)
&=&
\frac{t^2 \left(q^2+t q+t^2\right) \left(q^5+t^4 q^4+t^2 q^3+t^3 q^2+t
   q+t^5\right)}{\left(q^3-1\right)^2
   \left(q^3+1\right) \left(t^3-1\right)^2 \left(t^3+1\right)}
\nonumber\\
&=&
\frac{1}{2}h_3(t_1^2, t_2^2)
+
\frac{1}{2} \left( h_3(t_1, t_2) \right)^2
\nonumber \\
&=&
\mbox{Coeff}_{\nu^2}\left(
PE\left( h_k(t_1, t_2)\right)
\right).
\end{eqnarray}

\renewcommand{\theequation}{\Alph{section}.\arabic{equation}}
\appendix
\section{Partitions}
\label{sec;appendix}

A partition $\lambda$ 
is a nonincreasing sequence of non-negative integers: 
$\lambda=(\lambda_1,\lambda_2,\cdots)$. 
Partitions are often identified with the Young diagrams. 
Take a box $s=(i,j) \in \lambda$ and let $H_s$ be the hook. 
\begin{eqnarray}
H_{s=(i,j)}
=
\{(k,l) \in\lambda|
k=i,l\geq j,\,\,
\mbox{or}\,\,\,
l=j, k>i
\}.
\end{eqnarray}
The arm length, the leg length and the hook length are 
\begin{eqnarray}
a(s)=\lambda_i-j,
& \hspace{10mm}
l(s)=\lambda^t_{j}-i, 
& \hspace{10mm}
h(s)=a(s)+l(s)+1\,,
\label{eq;arm and leg}
\end{eqnarray}
where 
$\lambda^t =(\lambda^t_1, \lambda^t_{2}\cdots)$ 
is the dual partition. 
The hook $H_s$ is called $r$-hook if $h(s)=r$.

Fix a positive integer $k$. 
A partition $\lambda$ is called $k$-core 
if it does not contain any $k$-hook. 
Otherwise, 
another partition $\lambda'$ is obtained 
from $\lambda$ by removing one of the $k$-hooks. 
The procedure can be continued until 
we obtain a $k$-core. 
This partition turns out to be independent from 
the way $k$-hooks are removed and 
is denoted by $\lambda_{\mathrm{core}}$. 
The set of removed $k$-hooks constitutes 
$k$-tuple of partitions 
$\lambda^{(1)},\cdots,\lambda^{(k)}$. 
These partitions are obtained uniquely from $\lambda$ 
and are called $k$-quotients. 
In the case of $\lambda_{\mathrm{core}}=\phi$, 
we can read the $k$-quotients 
$\lambda^{(1)},\cdots,\lambda^{(k)}$ 
from the following relation: 
\begin{eqnarray}
\left\{\lambda_i-i ; i\geq 1
\right\}
=
\bigcup_{r=1}^{k}
\left\{
k(\lambda_{i_r}^{(r)}-i_r)+r-1; i_r \geq 1
\right\}\,.
\label{eq;N to 1 map}
\end{eqnarray}

\subsection*{Acknowledgements}

We thank to A.~Hanany
 for useful discussions
 and encouraging us.
Y.~N. is supported in part
 by JSPS Research Fellowships
 for Young Scientists. 
T.~N. is supported in part by 
Grant-in-Aid for Scientific Research 15540273.


\end{document}